 \documentclass[aps,pra, twocolumn,groupedaddress,showpacs] {revtex4-1} 

 \usepackage{amssymb} 
 
 \usepackage{geometry}
\usepackage{graphicx}
\usepackage{epstopdf}

\def\AOM {acousto-optic modulator}

\def\FVC {frequency-to-voltage converter}

\def\hf {hyperfine}

\def\LIA {lock-in amplifier}

\def\ppKTP {periodically poled potassium titanyl phosphate}

\def\PMT {photo-multiplier tube}

\def\rf {repetition frequency}

\def\sa {saturated absorption}

\def\snr {signal-to-noise ratio}

\def\wrt {with respect to}

\def\ARC {Australian Research Council} 
\def\EQUS {Engineered Quantum Systems}
\def\UWA {University of Western Australia}

\newcommand{\degrees}{$^{\circ}$}
\newcommand{\degC}{$^{\circ}$C}

\newcommand{\uK}{$\mu$K}  
\newcommand{\uW}{$\mu$W}

\newcommand{\coolingT}{$^{1}S_{0}-\,^{3}P_{1}$}  
\newcommand{\coolingTfull}{$(6s^{2})\,^{1}S_{0} -(6s6p)\,^{3}P_{1}$}

\newcommand{\clockT}{$^{1}S_{0}-\,^{3}P_{0}$}

\newcommand{\si}{$\sim$}

\newcommand{\Ybzero}{$^{170}$Yb}
\newcommand{\Yb}{$^{171}$Yb}	
\newcommand{\Ybtwo}{$^{172}$Yb}
\newcommand{\Ybthree}{$^{173}$Yb}
\newcommand{\Ybfour}{$^{174}$Yb}
\newcommand{\Ybsix}{$^{176}$Yb}

\begin{document}

\title{Hyperfine constants and line separations for the  $^{1}S_{0}-\,^{3}P_{1}$ intercombination line in neutral ytterbium with sub-Doppler resolution} %


\author{ P.~E.~Atkinson, J.~S.~Schelfhout  and J.~J.~McFerran }   
\email[]{john.mcferran@uwa.edu.au}  

\affiliation{ARC Centre of Excellence for Engineered Quantum Systems, Department of Physics, University of Western Australia, 6009 Crawley, Australia}  


\date{\today}

\begin{abstract}

Optical frequency measurements of the intercombination line $(6s^{2})\,^{1}S_{0} -(6s6p)\,^{3}P_{1}$ in the isotopes of ytterbium are carried out with the use of  sub-Doppler fluorescence spectroscopy  on an atomic beam.  A dispersive signal is generated to which a master laser is locked, while frequency counting  of an auxiliary beat signal is performed via a frequency comb referenced to a hydrogen maser. 
The relative separations between the lines are used to evaluate the $^{3}P_{1}$-level magnetic dipole and electric quadrupole constants for the fermionic isotopes. The center of gravity for  the $^3P_1$ levels in $^{171}$Yb and $^{173}$Yb are also evaluated, where we find significant disagreement with previously reported values.   These hyperfine constants provide a valuable litmus test for atomic many-body computations in ytterbium. 
\end{abstract}


                           
                            
                            \pacs{32.10.Fn; 32.30.-r;  42.62.Eh;   42.62.Fi}  
                            



\maketitle 

\vspace{0.5cm}

\section{Introduction}

Neutral ytterbium is exploited in a range of atomic physics experiments.  In  ultracold gases, studies  include, Bose-Einstein condensates in lattices~\cite{Tai2015, Yam2016, Bou2017},  degenerate Fermi gases~\cite{Oza2018, Nak2016, Han2011,Dor2013},  artificial gauge potentials~\cite{Dal2011,Tai2012, Sca2014}, quantum many-body simulations~\cite{Bra2013b,Hof2015}, 
and ultracold molecules~\cite{Gut2018, Tai2016}.  In parallel, ytterbium has been used to develop one of the world's most accurate atomic frequency references~\cite{McG2018, Sch2017, Piz2017,Tak2015,Nem2016}. 
Ytterbium gained attention in the investigation of atomic parity nonconservation~\cite{Tsi2009, Ant2017} when the level of violation was shown to be about one hundred times stronger than in Cs~\cite{Woo1997}.  Such experiments examine  nuclear physics at low energy and 
  are a means to explore physical behaviour beyond the Standard Model of elementary particles~\cite{Gin2004, Dzu2012, Saf2018, Ant2019}. 
The comparison between measured and computed hyperfine structure constants acts as an important test for the modelling of atomic wave functions in the nuclear region~\cite{Dzu2002a,Dzu2011}, and provides information  for atomic many-body calculations relevant to atomic parity violation and permanent electric dipole moments~\cite{Dil2001, Dzu2007, Der2008, Cam2016}. 
 Furthermore, knowledge of the hyperfine constants is applied to the analysis of   photoassociation spectra in ultracold  ytterbium~\cite{Han2018a}.  

%
Given the significance of ytterbium and the importance of its hyperfine constants, we have applied a 
 sub-Doppler spectroscopy technique to   measure the optical frequencies of the  intercombination line (ICL) in the isotopes of ytterbium.    This grants approximately a ten-fold reduction in the width of spectral features compared to previous measurements. 
From the optical frequencies   we deduce the various line separations, and for  the fermionic isotopes, \Yb\ and \Ybthree, we evaluate the hyperfine constants;  i.e., the magnetic dipole constants, $A(^3P_1)$ for \Yb\ and \Ybthree, and the electric quadrupole constant, $B(^3P_1)$ for \Ybthree.   We find good agreement with the earlier work by Pandey \textit{et al.}~\cite{Pan2009} for the $A$ coefficients and reasonable agreement for $B(^3P_1)$. However, we find significant disagreement for the center of gravity values.  With our new  $A$ constants, we compute an updated value of the hyperfine anomaly for the $(6s6p)$ $^3P_1$ state.   We note that Doppler-free absorption spectroscopy has been applied to other lines in neutral Yb~\cite{Ber1992a} and to ionised ytterbium to extract hyperfine constants~\cite{Ber1992a, Maa1994} (where the resolution was lower). 

\section{Experiment}  

\begin{figure}[h]			
 \begin{center}
{		
  \includegraphics[width=7.5cm,keepaspectratio=true]{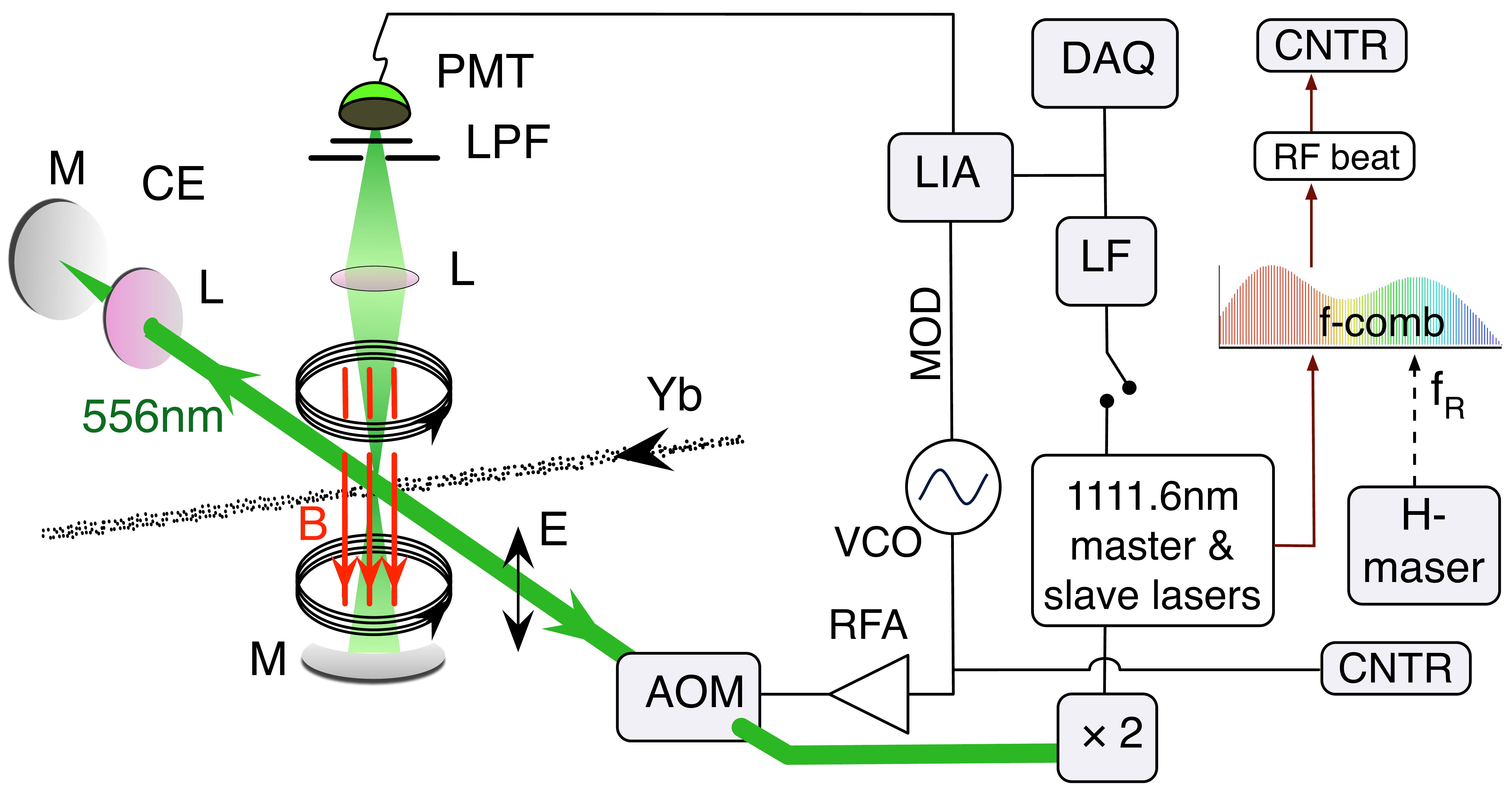}}   
\caption[]{\footnotesize     
 A sketch of the setup for saturated absorption spectroscopy of the \coolingTfull\ transition on an atomic beam of Yb.   The \LIA\ (LIA) outputs an error signal which is fed back to the 1112\,nm laser for stabilization to the intercombination line.   This light is also used to generate a heterodyne beat note with an element of a frequency comb, which is frequency counted.  The magnetic ($B$) field is applied to cancel the Earth's vertical $B$-field component. 
 AOM, \AOM; CNTR, frequency counter; CE, cat's eye reflector;  DAQ, data acquisition for spectra; $E$, electric field polarization; f-comb, frequency comb (in the near-IR); $f_{R}$, repetition frequency; L, lens;  LF, loop filter; LPF, long-wavelength pass filter (505\,nm); 
 M, mirror;  MOD, modulation; PMT, \PMT; RFA, radio frequency amplifier; VCO, voltage controlled oscillator. 
  }
   \label{SASexp}  %
\end{center}
\end{figure}

A sub-Doppler spectroscopy scheme that relies on saturated absorption  
is used to generate spectra of the individual ICL transitions.  The method is outlined in Fig.~\ref{SASexp}.  Ytterbium atoms effuse through narrow collimation tubes extending from an oven held at 450\degC\ and under vacuum.   A 556\,nm laser beam is retroreflected with a mirror-lens combination (cat's eye) with its optical axis orientated in the horizontal plane. The focal length of the lens is 75\,mm and the optics   are set in  a Thorlabs cage system.  The green beams and atomic beam are made orthogonal by centering the saturated absorption signal (Lamb dip) on the Doppler background, when applying first order detection. Fluorescence is detected  by way of a photomultiplier tube (PMT, Hamamatsu H10492-003) that is located above the atom interrogation zone.  There is a concave, silver coated mirror located below the interaction zone  with a focal length of 40\,mm, matching the distance from the atoms.   A 30\,mm focal length convex lens above the atoms directs fluorescence  onto the photomultiplier cell. The atomic flow rate, as determined by the DC level of the PMT, is \si\,$3\times10^{10}$\,s$^{-1}$.
 Detection on the third harmonic is performed for frequency measurements to minimize the influence of the Doppler background.     Helmholtz coils are used to create a vertical bias field to null the  vertical component of the Earth's magnetic field and to avoid any Zeeman splitting.     Frequency modulation at 33\,kHz  is applied to the 556\,nm light by means of an \AOM\ (AOM).   The modulation is sourced from a \LIA\ (LIA) (SRS SR830)  that receives the PMT signal and outputs the third harmonic dispersive signal. 
Generation of the 556\,nm light is described in~\cite{Sal2017}.  In brief, a fiber laser  
 injection-locks a 50\,mW ridge waveguide diode laser 
 for amplification. This light is frequency doubled in a resonant cavity containing a \ppKTP\ crystal with 40\% conversion efficiency~\cite{Kos2015}. Only  $\sim200$\,\uW\ of the light is used for the saturation spectroscopy.

%

To scan across the resonance lines we lock the 1112\,nm master laser to a mode (tooth) of a frequency comb (Menlo Systems FC1500) and sweep the green light's frequency by use of the AOM.   The frequency comb is indirectly locked to a hydrogen maser.   For locking the  1112\,nm laser to the comb, an optical beat is generated between the 1112\,nm light and a comb tooth, producing an RF beat signal that is frequency divided, passed through a frequency-to-voltage converter (FVC), summed with a voltage reference, then sent to a PI  filter.  The output of which is fed back to the 1112\,nm laser via a high voltage driver and piezo transducer~\cite{McF2018}.  Further details are given in Sect.~\ref{Sect3}.  

\begin{figure*}[] 
 \begin{center}
{		
  \includegraphics[width=16cm,keepaspectratio=true]{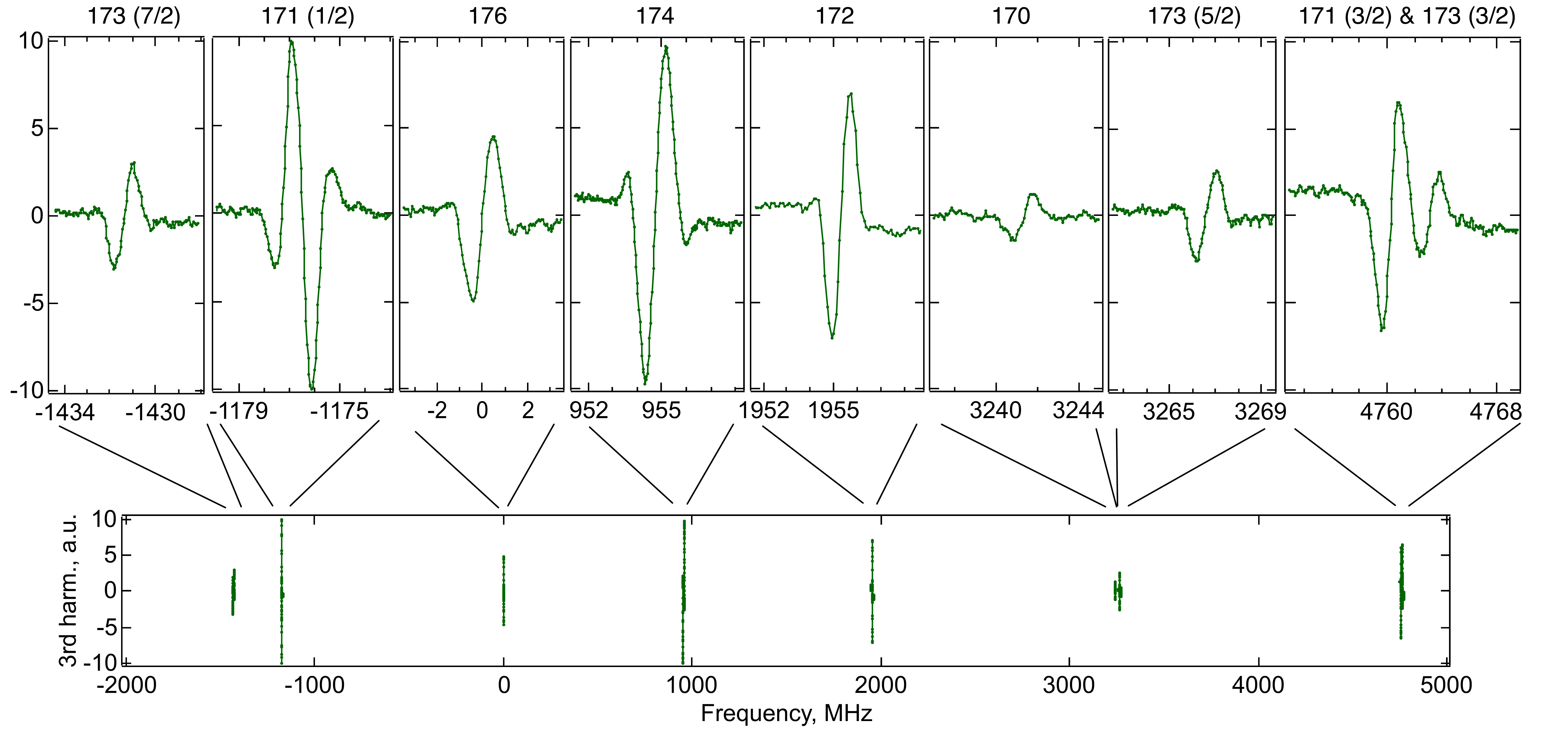}}   %
\caption[]{\footnotesize     
  Sub-Doppler spectra of the intercombination line in ytterbium for all the abundant isotopes, displaying the relative strengths of the various lines.  Each spectrum is labelled by the atomic number along with the upper $F$ state in parentheses for the odd isotopes. All traces are saturated absorption signals except  for \Yb\ ($F'=1/2$), which  is an inverted crossover resonance. $a.u$., arbitrary units. } 
   \label{SpectraAll}  
\end{center}
\end{figure*}

\begin{figure}[h]			
 \begin{center}
{		
  \includegraphics[width=8cm,keepaspectratio=true]{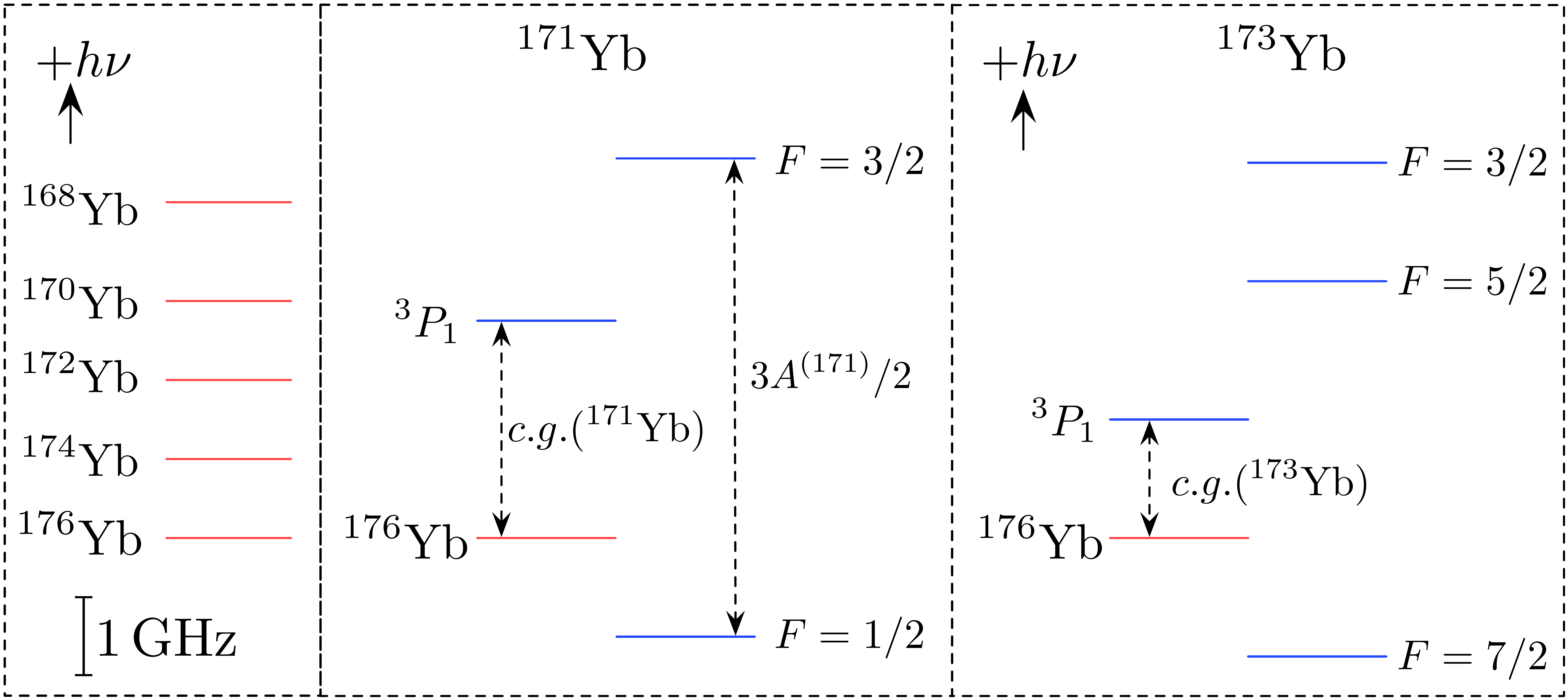}}   %
\caption[]{\footnotesize     
Frequency separation and sequence of the ytterbium isotope levels, and hyperfine level structure of \Yb\ and \Ybthree\ drawn to scale (note the inversion \wrt\ $F$ for \Ybthree).  The shift of the hyperfine levels in terms of the hyperfine constants is given in Table~\ref{HFSummary3} of the appendix.  }  
   \label{HFlevels}  
\end{center}
\end{figure}

 Spectra for all the   lines except $^{168}$Yb (with about one twentieth the abundance of \Ybzero) are shown in  Fig.~\ref{SpectraAll}.
 For comparison, we include a scaled energy level diagram  of the isotope shifts and hyperfine splittings in Fig.~\ref{HFlevels}.
 The spectral widths of Fig.~\ref{SpectraAll}  are not lifetime (184\,kHz) or transit time (70\,kHz) limited, rather,  they are broadened due to the applied modulation~\cite{Sal2017}  and the optical intensity of the probe light.   
  The intensity is \si70\,$I_\mathrm{sat}$,  where  $I_\mathrm{sat}=1.4$\,W$\cdot$m$^{-2}$ is the saturation intensity for the \coolingT\ transition.  The intensity and modulation amplitude (1.2\,MHz) were chosen to provide sufficient \snr\ (SNR) for robust servo operation, in particular for the weaker lines.  The sweep rate for the spectra in  Fig.~\ref{SpectraAll} was 260\,kHz per second.
  Integration of the traces  (once or three times) produces the familiar Lamb dip.
  The center of the discriminator acts as the lock point for a feedback loop when frequency measurements are undertaken.  Under this condition the FVC servo is no longer engaged and the atomic resonance signal is used to create a correction signal that is, instead, sent to the piezo element in the 1112\,nm master laser  via a high-voltage driver. 
   The right-most spectrum  of  Fig.~\ref{SpectraAll} shows the discriminating slopes for two closely separated transitions:  the \Yb\ ($F'=3/2$) and \Ybthree\  ($F'=3/2$) lines; the former being the stronger of the two.   The resolution of these two lines at zero $B$-field has not been previously been presented. 
 It enables a direct measurement of the frequency difference between the two and is relevant to the determination of the \hf\ constants in Sect.~\ref{HFConstants}.

 The spectrum for \Yb\ ($F'=1/2$) shown in Fig.~\ref{SpectraAll} is not a saturated absorption signal, but is instead an inverted crossover resonance between two Zeeman transitions~\cite{Sal2017}, hence has a central slope that is inverse to the others.  It is generated when a bias magnetic  field is applied, in this case 0.1\,mT.  It has been included to show the relative strength of this resonance to the saturated absorption signals.   The center of this resonance coincides with the center of the \Yb\ ($F'=1/2$)  line, as shown in \cite{Sal2017}.

To record  the center-frequencies of the transitions the optical beat  between the sub-harmonic light at 1112\,nm light and the comb tooth is frequency counted  when the 1112\,nm laser is locked to the center of the dispersive curve.  
After an averaging time of 30\,s the Allan deviation is usually between 1\,kHz and 5\,kHz, depending on the SNR of the signal. 
Evaluation of the optical frequency is made through, 

\begin{equation}
\label{ }
\nu = 2 (n f_r + f_o \pm f_b) + f_\mathrm{aom}
\end{equation}
where $f_r$ is the comb's repetition rate, $f_o$ its carrier-envelope offset frequency (maintained at $-20$ MHz), $f_b$ is the measured beat frequency,  $n$ is the mode number of the comb (which differs for many of the lines), and  $f_\mathrm{aom}$ is the RF drive frequency of the \AOM\ that is in the path of the 556\,nm light.
The comb's repetition rate is controlled by mixing its fourth harmonic with a 
signal that is the frequency sum of a direct digital synthesizer  (DDS, \si20\,MHz) and a dielectric resonator oscillator (DRO) fixed at 980\,MHz.  The DRO is phase locked to a 10\,MHz signal from the hydrogen maser, as is the DDS.   The \rf\ is simply $f_r = (f_\mathrm{DDS} +f_\mathrm{DRO})/4$.

The beat signal between the comb and the 1112\,nm light is band-pass  filtered  at 30\,MHz.  We maintain the use of this filter (and approximate beat frequency) by changing the comb's repetition rate across the different lines by changing $n$ and  adjusting the DDS frequency.   These parameters are listed in Table~\ref{SystShifts} of the appendix. 

\section{Isotopic shifts, hyperfine separations and systematics}  \label{Sect3}

We have identified the important systematic shifts associated with the frequency measurements.   Since many of the shifts are common to all lines, they tend to be heavily suppressed when it comes to frequency differences between the lines.    
The dominant shift arises from beam alignment, and there is  
 also an uncertainty associated with identifying  the line-center, which over several  measurements presents itself as a statistical uncertainty.

\begin{figure}[h]			
 \begin{center}
{		
  \includegraphics[width=8.5cm,keepaspectratio=true]{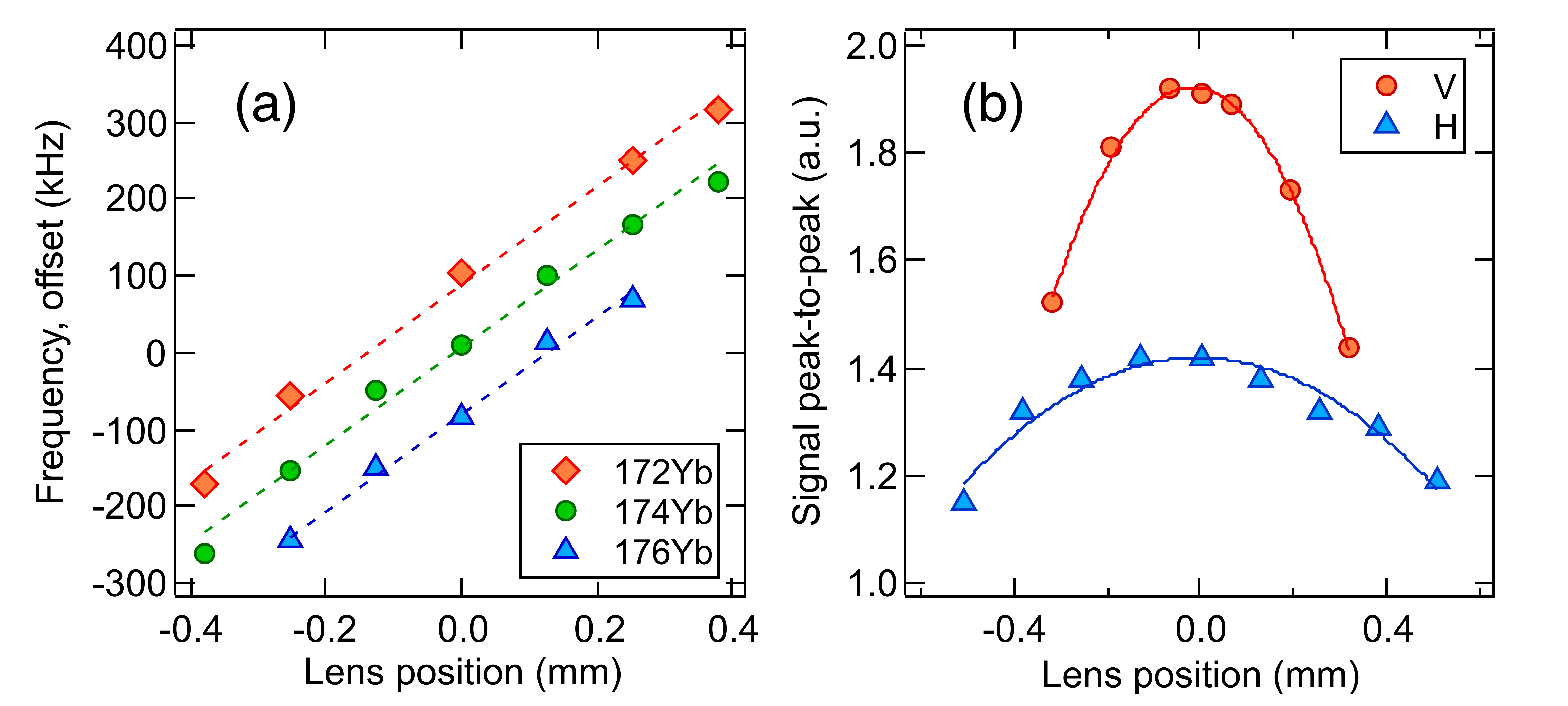}}   %
\caption[]{\footnotesize     
(a)  Frequency dependence on the  lens's vertical position in the cat's eye for  three bosonic ytterbium isotopes. Absolute optical frequencies have been subtracted and deliberate offsets have been applied  to the \Ybtwo\ and \Ybsix\ data  to more clearly show the slopes.  
  (b)  Signal strength versus  vertical (V) and horizontal (H) lens position.  }

   \label{Lensshift}  %
\end{center}
\end{figure}

To test the sensitivity to beam alignment, the ICL's optical frequency  was measured as a function of the position of the lens in the cat's eye that retroreflects the 556\,nm beam.
Fig.~\ref{Lensshift}(a)  shows the frequency shift versus lens position in the vertical direction (i.e. across the atomic beam) for \Ybtwo, \Ybfour\ and \Ybsix.  The gradients are very similar to one another at $630\,(30)$\,kHz\,mm$^{-1}$.    Measurements for horizontal lens displacements show much smaller sensitivities at $50\,(9)$\,kHz\,mm$^{-1}$.   
Relevant to the isotopic data is the relative shift between the isotopes, which is much smaller than the uncertainties of the slopes from the line fits. 
We are not aware of a model  that predicts the shift associated with the lens displacement in the vertical direction, but we know it to be dependent on the beam diameter in this direction. A larger spot size reduces the shift  inversely to the diameter. The 556\,nm beam profile is elliptical with $e^{-2}$ intensity  radii of 1.1\,mm and  2.0\,mm, where the major axis is aligned with the flow of the atoms, and has  
 beam divergences of 0.21\,mrad  and 0.10\,mrad for the respective dimensions (very close to the diffraction limited divergence).     We also varied the beam's  horizontal divergence from 0.1\,mrad to 0.4\,mrad and observed no frequency shift at a resolution of 3\,kHz.  Other reports show that beam size does not influence the Lamb dip frequency at the $\sim10$\,kHz level (for sufficiently low intensity)~\cite{Cou2005}.%

Although the relative shift between the bosonic isotopes cancels for lens displacement in the vertical direction, we can identify an optimum position to set the lens. This is done by observing the signal strength of the third harmonic signal for different lens positions.  Two examples are shown in Fig.~\ref{Lensshift}(b), where the upper (lower) trace is for vertical (horizontal) lens displacements.   From this we place an uncertainty on the lens position to be $\pm0.1$mm  and  $\pm0.2$mm for the vertical and horizontal directions, respectively.  These are relevant for absolute frequency measurements and for comparisons with some fermionic lines.

\begin{figure}[h]			
 \begin{center}
{		
  \includegraphics[width=8.8cm,keepaspectratio=true]{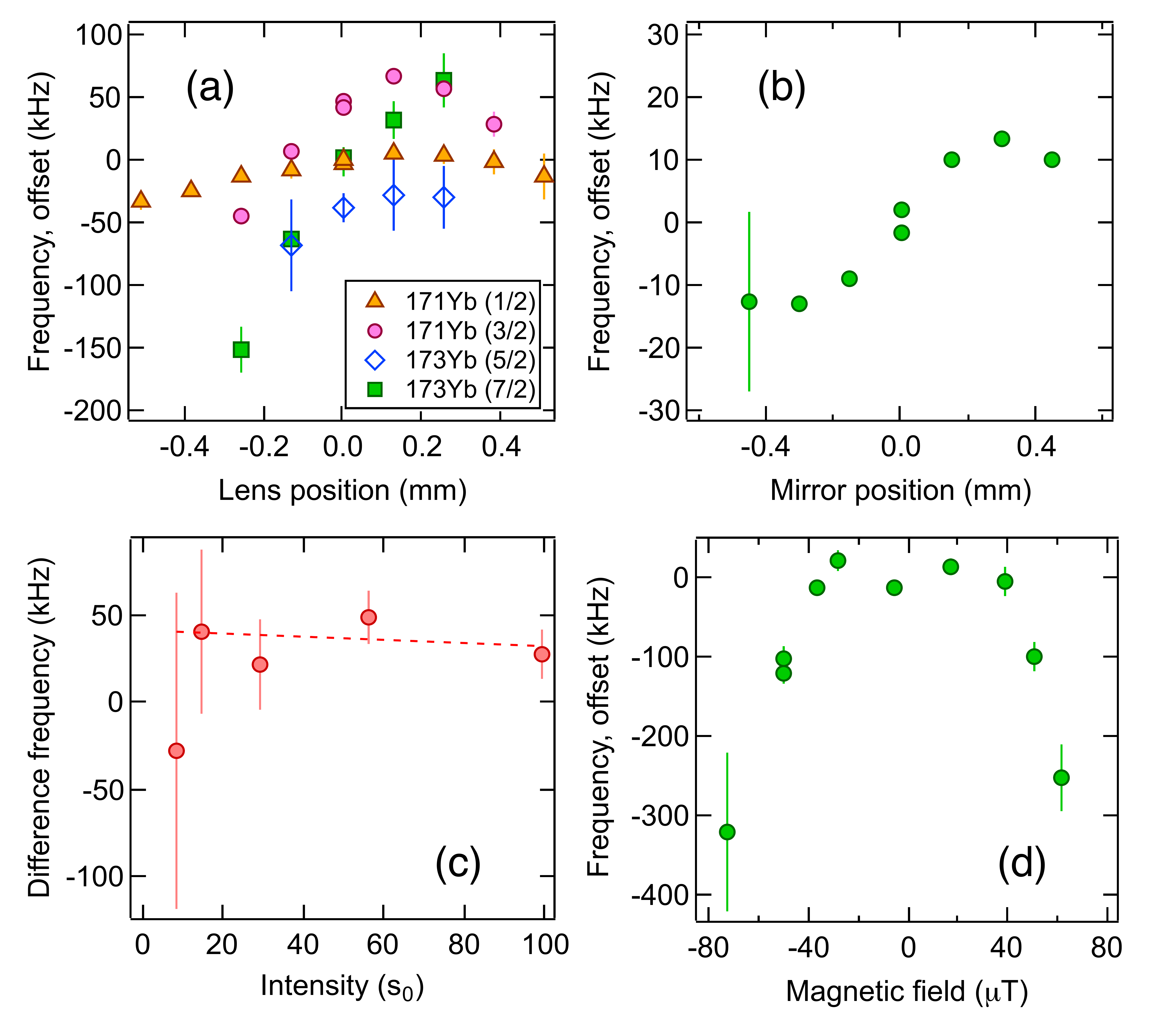}}   %
\caption[]{\footnotesize     
(a)  Absolute optical frequency dependence on the  lens's vertical position in the cat's eye for  several lines between hyperfine states.  Only the upper $F$ state is stated on the plot.  Some intentional vertical offsets have been applied. 
(b) Optical frequency of the  \Ybtwo\  intercombination line as a function of lens-mirror separation in the cat's eye.  The error bar is the approximate uncertainty (Allan deviation) on all the data points.
(c)  Frequency difference between  the \Yb\,(F'=3/2) and \Ybtwo\ lines as a function of 556\,nm beam intensity.  The frequency separation has been offset by  $2.80465$\,GHz. 
(d) Optical frequency of the  \Ybtwo\  line as a function of the vertical magnetic field.  The frequencies in (b) and (d\,) are offset by 539\,387\,600\,920\,kHz. 
}
   \label{Shifts2}  %
\end{center}
\end{figure}

The frequency change with lens displacement for the fermionic isotopes tends to be much less and also exhibits more curvature.  Examples are shown in Fig.~\ref{Shifts2}(a).  The energy level structure of  the odd isotopes is more complex than that of the even isotopes.  It is likely that the difference in behaviour is related to this.  
 For the frequency separations discussed below (and in Table~\ref{isoshifts}) the uncertainties for the fermionic levels are dominated by that associated with \Ybsix.  Likewise, the overall systematic uncertainties listed in Table~\ref{SystShifts} of the appendix are dominated by this systematic shift.

The lens-mirror separation in the cat's eye is set so that the reflected beam is parallel to the incident beam (more strictly, the wavefronts of the forward and reverse beams match each other)~\cite{Sny1975}.  We have measured the optical frequency of the  \Ybtwo\  intercombination line (ICL) as a function of lens-mirror separation.  The variation is not much greater than the resolution available here, as shown in Fig~\ref{Shifts2}(b).  There is also a point of inflection, which appears to match  the optimum mirror position. 
  We also expect any shift here to be common to all the lines and has negligible influence on the hyperfine constants we evaluate below. 
Ideally, when the cat's eye lens-mirror  separation is optimally set for retroflection, the center frequency of the saturation dip should not change with lens displacement in the horizontal direction~\cite{Mor1989,Cou2005}.  We find that changes in the  lens-mirror separation do not affect  the shift associated with the lens's horizontal position; i.e. the previously mentioned $50\,(9)$\,kHz\,mm$^{-1}$ shift cannot be completely nulled.  Its contribution is, at most, 10\,kHz in the uncertainty evaluations. 

The frequency difference between the \Yb\,(F'=3/2) and \Ybtwo\ lines as a function of 556\,nm beam intensity is shown in Fig.~\ref{Shifts2}(c\,).  The intensity was adjusted by use of the AOM in the path of the 556\,nm light.  The abscissa is expressed in terms of the saturation parameter, $s_0=I/I_\mathrm{sat}$. 
 There is no apparent light shift affecting the line separations within the uncertainty of the slope of the line fit.  We do observe a light shift of the absolute frequencies, but this shift is cancelled in a common mode fashion when evaluating the frequency separations.    The increased uncertainty at lower intensity arises due to the poorer SNR of the \sa\ signal.   In any particular measurement run all the lines are measured with the same power to within a few percent.   Across the measurement runs the variation was $\pm10$\,$s_0$ giving a shift uncertainty of 3\,kHz.   %

\begin{figure}[h]			
 \begin{center}
{		
  \includegraphics[width=8.8cm,keepaspectratio=true]{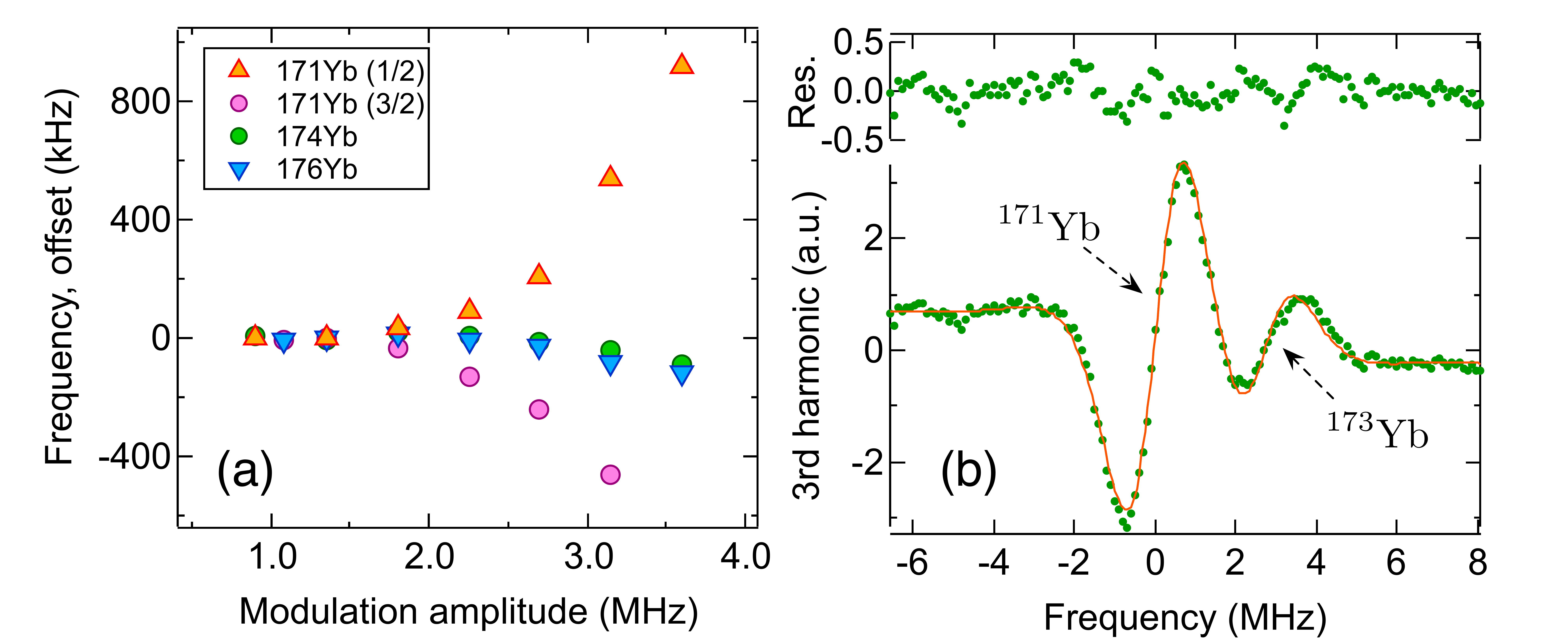}}   %
\caption[]{\footnotesize     
(a) Center-line frequency versus the modulation amplitude applied to the 556\,nm light.  Different optical frequencies have been subtracted for each isotopic line. (b) Saturated absorption spectrum for 
  \Yb\ ($F'=3/2$) and \Ybthree\  ($F'=3/2$).  The residuals of the curve fit are shown above.  
   }
   \label{SpectraDuo}  
\end{center}
\end{figure}

A significant systematic shift arises from the modulation amplitude that is applied to the 556\,nm light.  We summarise the effect in Fig.~\ref{SpectraDuo}(a), where the centerline frequency is plotted against the modulation amplitude.  Different optical frequency offsets have been applied to each data set to present the comparison between the isotopes.   At less that 1.5\,MHz amplitude the shift is insignificant, but at higher strengths the shift is sizeable, moreover, the shift is different between the isotopes; e.g. the size of the shifts for the \Yb\ lines is greater than that for the bosonic lines.   A related effect is described in \cite{Bur2004b}.  The inverse shift exhibited by \Yb\ ($F'=1/2$) is most likely due to it being an inverted dip (with 180\degrees\ phase shift).  Importantly, for all our isotopic and hyperfine frequency shift measurements,  we operate with a modulation where the shift is negligible (1.2\,MHz).  

There is a $-0.31$\,kHz second order Doppler shift~\cite{Bar1981} with each ICL, which may be considered constant across all the lines. The shift is calculated according to $\Delta\nu =-(\nu_0/2)(v^2/c^2)$, where $\nu_0$ is the optical frequency.  For the velocity we take the most probable velocity for an atomic velocity distribution produced by a thin tube, $v\approx (3k_B T/m)^{1/2}$, where $k_B$ is Boltzmann's constant, $m$ is the atomic mass, and $T$ the temperature (455\degC).  
There is also a recoil shift, $h\nu^2/(2mc^2)=3.7$\,kHz, but again, at our resolution, this is the same across all the lines. 

As described above, we use a bias magnetic field to cancel the vertical component of the Earth's magnetic field.   There are several methods  we can apply to find the required bias current.   One is by examining the strength of the inverted crossover resonances (ICRs) for opposite field settings,  as described in~\cite{Sal2017} (e.g. for \Yb, $F'=1/2$).  Another is by measuring the  Zeeman splitting of the  bosonic isotope lines when $\sigma^{\pm}$ excitation is permitted (e.g. by setting the light polarization perpendicular to the magnetic field direction).    In the  $\pi$-polarization  configuration the strength of the \sa\ signal is strongest at zero $B$-field.  We can cancel the Earth's field to within $\pm0.006$\,mT  ($\pm0.06$\,G).    Over a range of $\pm0.04$\,mT ($\pm0.4$\,G) there is no observable change in the line-center frequencies, as seen in Fig.~\ref{Shifts2}(d).  Beyond this range the distorted line shape influences the frequency measurements (and is not a Zeeman shift).  %

 \begin{figure}[h]			
 \begin{center}
{		
  \includegraphics[width=8.4cm,keepaspectratio=true]{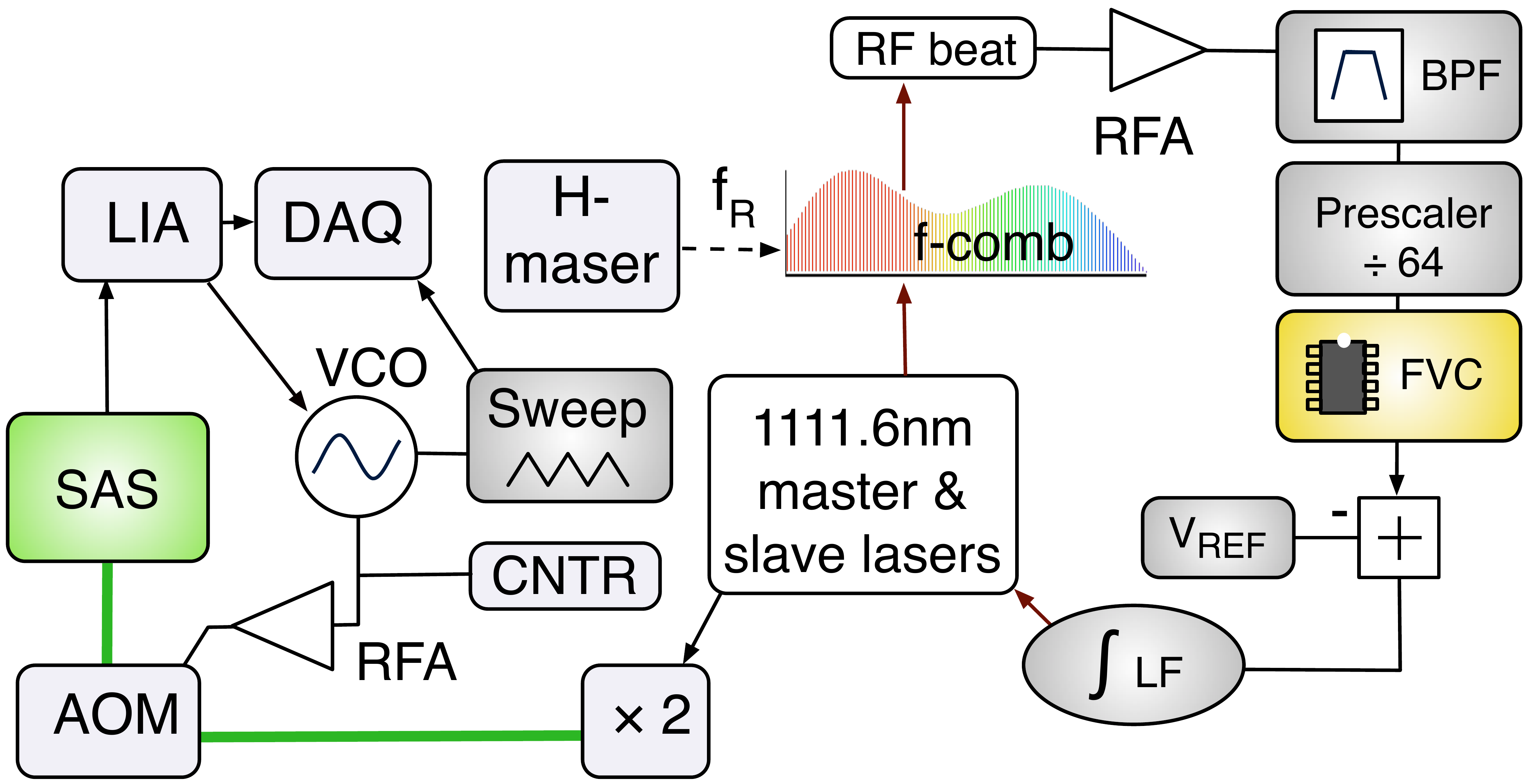}}   %
\caption[]{\footnotesize     
Experimental setup for generating the calibrated line spectra.  A servo using a \FVC\ (FVC)  locks the 1112\,nm laser to an element of the frequency comb.  A function generator is used to sweep the green light's frequency by use of an \AOM\ (AOM).  BPF, band pass filter;  CNTR, frequency counter;  DAQ, data acquisition; $f_{R}$, repetition frequency;  LF, loop filter;
  SAS, saturated absorption scheme.   
}
   \label{Linescanning}  
\end{center}
\end{figure}

The uncertainty associated with locking to the center of the line we evaluate as  $\Delta f\approx \Delta s/\mathrm{SNR}$, where $ \Delta s$ is the width of the discriminator signal, which 
 is approximately 900\,kHz (governed  mostly by the applied modulation strength). The SNR varies from  \si10 for \Ybzero\ to \si80 for \Ybfour\ in a measurement time of 300\,ms.    The line centering is set by adjusting the output offset of the LIA, but in most cases this  is zero (to the resolution of the LIA).  
The stronger isotopic lines  dictate the line-centering uncertainty for other lines.    Over many measurements this shift becomes randomized, so its inclusion along with later statistical uncertainties creates a conservative estimate on the error bounds.  

To measure the frequency difference between the \Yb\ ($F'=3/2$)   and \Ybthree\ ($F'=3/2$)  lines, the  experimental  scheme shown in Fig.~\ref{Linescanning} was used.   A function generator sweeps the probe laser's frequency via the AOM and calibration is made by recording the AOM's frequency.
  The LIA signal and  sweep signal are recorded simultaneously with a multichannel data acquisition unit (Agilent 34970A). Twenty spectra have been recorded and  curve fitting  applied to extract the separation.   The fitting function takes the form~\cite{Dem2008},
\begin{equation}
\label{ }
P^{(3)}(\nu) =  c_1 P^{(3)}_{171}(\nu) + c_2 P^{(3)}_{173}(\nu+\Delta \nu),
\end{equation}
with
\begin{equation}
\label{ }
P^{(3)}_A(\nu') =  \frac{(\nu'-\nu_0)[(\nu'-\nu_0)^2 - (\gamma/2)^2]}{[(\nu'-\nu_0)^2 +(\gamma/2)^2]^4},
\end{equation}
where $\nu_0$ is the center-frequency of the \Yb\ ($F'=3/2$) line, $\Delta \nu$ is the separation between the line centers, $\gamma$ is the line width (in hertz), and $c_1$ and $c_2$ are scale factors for the amplitudes. 
From the series of line shape  fits we find the separation to be 2.679(24)\,MHz in close accord with Pandey \textit{et al.}~\cite{Pan2009}.  An example of the data and curve fit are seen in Fig.~\ref{SpectraDuo}(b)  (a background slope is  applied to the curve fitting, which has little influence on the line separations).   A search for  systematics shifts did not reveal anything comparable to the statistical variations.


 A summary of our line measurements are presented in Table~\ref{isoshifts}.  Each entry is a mean of seven recent measurements, together with six measurements recorded in 2016.  The frequencies of the individual lines are recorded in  succession in one measurement run (lasting several hours), and each run is separated by a week.   The uncertainty for each line separation is an rms combination of statistical and systematic uncertainties, and the standard deviation of the sample mean is assumed for the statistical uncertainty.   The mean optical frequency for \Ybsix\ across the measurements is 539\,385\,645\,457\,(87)\,kHz, where the quoted uncertainty here is simply the sample standard deviation.       While absolute frequency measurements are not the focus of this work, we regard this is  an accurate measurement, since our own measurements of the \clockT\ clock-line in Yb agree to within 10\,kHz of the accepted value.  Our clock-line measurements were carried out on a atomic cloud sample at a temperature of $\sim30$\,\uK.  A less accurate experimental method in relation to this is described in ~\cite{Nen2016}.  

 \begin{table}
 \caption{  Summary of the measured \coolingT\ isotopic shifts and hyperfine splittings in ytterbium.
Some previously reported values are included for comparison.  For our data the quoted uncertainty is the rms sum of systematic and statistical uncertainties.    \label{isoshifts}} 
 \begin{ruledtabular}
 \begin{tabular}{llll}
 &  \hfill Shift from  & $^{176}$Yb (MHz) & \\ 
   \vspace{-0.3cm}  \\
  \hline
 Transition       & This work 	& Ref.~\cite{Pan2009} & Ref.~\cite{Wij1994} \\ 
 \hline
$^{173}$Yb\,($\frac{5}{2}\rightarrow\frac{7}{2}$)  & -1431.392\,(35) & -1431.872\,(60) & -1432.6\,(12)  \\ 
$^{171}$Yb\,($\frac{1}{2}\rightarrow\frac{1}{2}$)  & -1176.412\,(87)   & -1177.231\,(60)  & -1177.3\,(11) \\ 
$^{176}$Yb & 0  &  0 & 0  \\ 
$^{174}$Yb & 954.734\,(31) & 954.832\,(60) &954.2\,(9) \\ 
$^{172}$Yb	& 1955.526\,(36) &  1954.852\,(60)  &1954.8\,(16) \\ 
$^{170}$Yb	& 3241.342\,(43) & 3241.177\,(60) & 3241.5\,(28)  \\ 
$^{173}$Yb\,($\frac{5}{2}\rightarrow\frac{5}{2}$) &  3266.557\,(95) & 3266.243\,(60) & 3267.1\,(28)  \\ 
$^{168}$Yb	&- & 4609.960\,(80) & 4611.9\,(44)  \\ 
$^{171}$Yb\,($\frac{1}{2}\rightarrow\frac{3}{2}$) & 4760.247\,(64) & 4759.440\,(80) & 4761.8\,(37) \\ 
$^{173}$Yb\,($\frac{5}{2}\rightarrow\frac{3}{2}$) & 4762.926\,(77) & 4762.110\,(120) & 4761.8\,(37) \\ 
 \end{tabular}
 \end{ruledtabular}
 \end{table}

  There are  three frequency measurements that have shifted beyond the error bounds between this work and our previous work, namely those for \Yb\ ($F'=1/2$ \& $F'=3/2$) and  \Ybthree\ ($F'=3/2$).  
 This is primarily  because during our first measurement campaign we used a higher modulation strength (\si3\,MHz) and as described above [Fig.~\ref{SpectraDuo}(a)], this creates a systematic shift.   
 It also broadens  the resonances, so in the case of \Yb\ ($F'=3/2$),  the adjacent \Ybthree\ ($F'=3/2$) line was not properly resolved and was influencing  the line shape.    
  We can now resolve these lines and  have  more accurate measurements of the separation between them. 
   The data from 2016 were not used in the mean values reported here for these three lines.

\begin{figure*}[]			
 \begin{center}
{		
  \includegraphics[width=17cm,keepaspectratio=true]{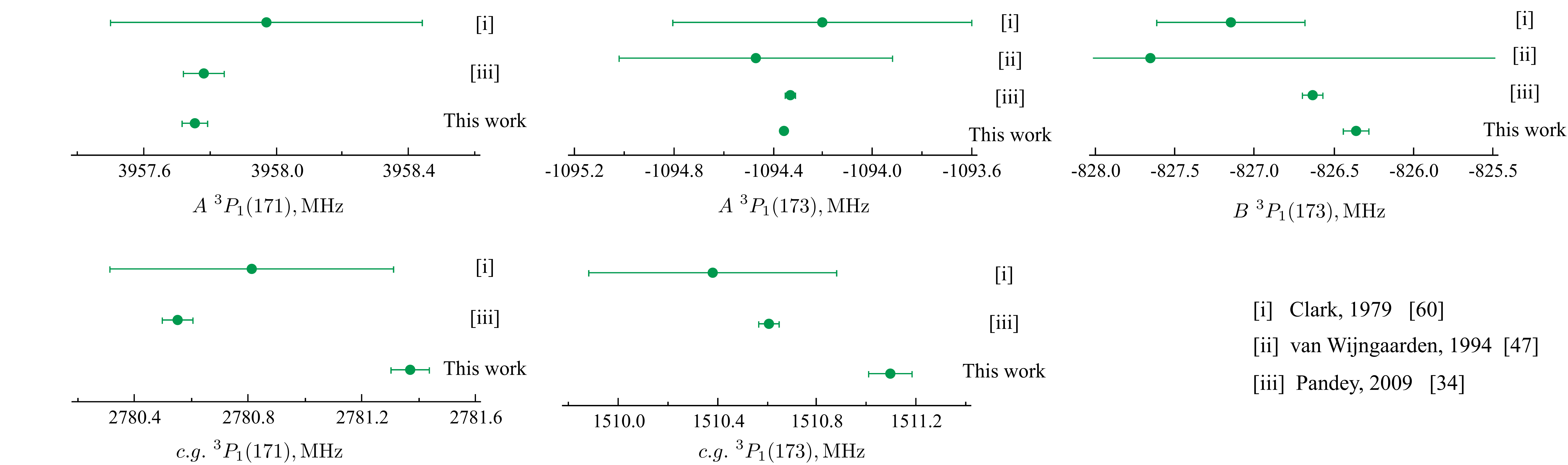}}   %
\caption[]{\footnotesize     
Summary of the hyperfine constants and centers of gravity for the intercombination line in ytterbium.  We make comparisons to some previous measurements.
 }
   \label{Constants}  %
\end{center}
\end{figure*}

 We have noticed that the statistical uncertainty for many of the fermionic lines is greater than that for the bosonic lines in Table~\ref{isoshifts}.  We attribute this to the systematic shift associated with the vertical alignment, as discussed earlier.  The shift cancels more strongly between the bosonic lines than it does between the \Ybsix\ line and the fermionic lines.   The hyperfine constants, described in the next section, only depend on the differences between the hyperfine levels, therefore the error analysis  incorporates uncertainties associated with the difference between hyperfine lines, not those \wrt\ \Ybsix.  
  Only in the case of the centers of gravity are the uncertainties relating to \Ybsix\ included. 
 The line separations and uncertainties pertaining to the 
 hyperfine constants are listed in Table~\ref{HFSummary1}.  For each odd isotope the table lists the pair of upper $F$ states, the corresponding frequency separation and both the systematic and statistical uncertainties. 

  \begin{table} [] \caption{Frequency differences relevant to the hyperfine constants along with systematic and statistical uncertainties.} \label{HFSummary1}
 \begin{ruledtabular}
 \begin{tabular}{clclclclc}
   \vspace{-0.3cm}  \\
Isotope & $F'$ pair &  $\Delta f$ & $\Delta u_\mathrm{stat}$  &  $\Delta u_\mathrm{syst} $\\  
  &  &  (MHz) &  (MHz) &  (MHz) \\  
  \vspace{-0.3cm}  \\
 \hline
$^{171}$Yb & ($\frac{1}{2},\frac{3}{2}$) & 5936.632 &  0.027 & 0.039 \\
$^{173}$Yb & ($\frac{3}{2},\frac{5}{2}$) & 1496.375 &  0.058 & 0.030 \\
$^{173}$Yb & ($\frac{3}{2},\frac{7}{2}$) & 6194.306 &  0.028 & 0.024 \\
$^{173}$Yb & ($\frac{5}{2},\frac{7}{2}$) & 4697.939 &  0.071 & 0.051 \\
 \end{tabular}
 \end{ruledtabular}
 \end{table}

In cases where the hyperfine lines are not clearly resolved, quantum interference may arise causing line-center shifts depending on the polarization of the probe light~\cite{Bro2013, Kle2016}.  All the ICL hyperfine lines in ytterbium are fully resolved apart from \Yb\ (3/2) and \Ybthree\ (3/2), which although resolved, do have portions of their line shapes that are shared.  We performed two tests to examine the effect of the light polarization.  In the first test the light polarization was  rotated 90\degrees\ and multiple spectra were recorded.  The signal strength fell to 75\% of that for the vertically polarised light. Using the curve fitting procedure described above, the mean separation was found to be 2.687(13)\,MHz across seven measurements $-$ only 8\,kHz  difference from the value above.   In a second test the frequency of the \Yb\ (3/2) line was measured for two orthogonal polarizations multiple times.  There was no frequency shift observed within the \si4\,kHz resolution of the measurements. 

\vspace{2cm}  
\section{Hyperfine constants}   \label{HFConstants}
 
The energy shift associated with  the magnetic dipole and electric quadrupole interactions is given by~\cite{Ari1977}  

\begin{equation}
\label{HFEq}
E_{HF} =\frac{1}{2}AhK + Bh\frac{3K(K+1)-4I(I+1)J(J+1)}{8I(2I-1)J(2J-1)},
\end{equation} 
where $K = F(F+1)-I(I+1) - J(J+1)$,  $A$ and $B$ are the
magnetic dipole and electric quadrupole constants, respectively, and $h$ is Planck's constant.  In ytterbium there are two isotopes with non-zero nuclear spin: $^{171}$Yb (spin $I=1/2$ ) and $^{173}$Yb (spin $I=5/2$); of these only the latter experiences an electric quadrupole shift.
For the upper (6$s$6$p$) $^3P_1$ state the total electronic spin is $J=1$ and 
  the hyperfine interaction  
 generates states with quantum numbers of $F=\{1/2, 3/2\}$ for \Yb, and  $F=\{3/2, 5/2, 7/2\}$ for \Ybthree.   The hyperfine energy shifts based on Eq.~\ref{HFEq} are summarised in Table~\ref{HFSummary3} of the appendix.    The hyperfine constants  
 and centers of gravity for the odd isotopes are determined from the measured line separations, and are summarised in Table~\ref{HFSummary2}.  The final uncertainties assume uncorrelated errors between  the individual contributions.  Our constants are compared with previous determinations in Fig.~\ref{Constants}.  For the $A$, $B$ constants the agreement is mostly within the one sigma uncertainty bounds.    However,  there is a marked difference between our measurements of the centers or gravity and that of Pandey \textit{et al.}~\cite{Pan2009}.  We do not have an explanation for this difference.  For our centers of gravity  the  systemic shifts associated with \Ybsix\ are included in the error analysis.  In the case of Ref.~\cite{Wij1994}, we calculated the hyperfine constants from  their measured line separations. 

 \begin{table} [] \caption{Hyperfine constants and centers of gravity for \Yb\ and \Ybthree.   } \label{HFSummary2}
 \begin{ruledtabular}
 \begin{tabular}{lllll}
   \vspace{-0.3cm}  \\
Isotope & &  This work & Ref.~\cite{Pan2009} \\  
  \vspace{-0.3cm}  \\
 \hline
$^{171}$Yb &  $A$ &  3957.754(34) &  3957.781(63) \\
  &  c.g. from $^{176}$Yb   &  2781.369(66) & 2780.550(56) \\
  & & & \\
$^{173}$Yb &  $A$  & -1094.361(11)  & -1094.328(19) \\
 &  $B$  &  -826.351(79) &-826.635(67) \\
  &  c.g. from $^{176}$Yb &  1511.129(88) & 1510.607(39)   \\
 \end{tabular}
 \end{ruledtabular}
 \end{table}

For a point-like nuclear magnetic dipole,  one expects the ratio of  the $A$ constants and the ratio of nuclear $g$-factors to be identical.   However, because of the extended nuclear magnetization  and charge distribution the equality does not hold, instead the relationship may be written as~\cite{Bud1970, But1984},

\begin{equation}
\label{ }
 \frac{A^{(171)}}{A^{(173)}}  = \frac{g_I^{(171)}}{g_I^{(173)}} (1+  \Delta_\mathrm{HFA} ),
\end{equation}
where $\Delta_\mathrm{HFA}$ is the hyperfine anomaly, $g_I$ is the nuclear $g$-factor, $g_I = \mu_I /\mu_N I$, and $\mu_I$ is the   magnetic moment of the nucleus. Relying on the magnetic moment values from~\cite{But1984, Ols1972}, the anomaly  for the $^3P_1$ state evaluates to $\Delta_\mathrm{HFA} = -0.3857(51)$\%, which is consistent with, but more precise than a recent tabulated value~\cite{Per2013}.   The uncertainty is dominated by that of the magnetic moment ratio.   

\section{Conclusion}
We have measured the optical frequencies of the intercombination line for all the abundant  ytterbium isotopes using sub-Doppler fluorescence spectroscopy on an atomic beam.  From these, we find the isotope shifts and hyperfine separations for the $(4f)^{14}$ $6s6p$ $^3P_1$ level.  Our measurements show reproducibility for most lines over a period of three years. We have determined the hyperfine constants  $A(^{171}$Yb, $^3P_1)$, $A(^{173}$Yb, $^3P_1)$ and $B(^{173}$Yb, $^3P_1)$ and the centers of gravity for the \Yb\ ($^3P_1$) and \Ybthree\ ($^3P_1$) levels \wrt\ \Ybsix. 
Our $A$ values are in good agreement with previous measurements.  Our $B$ value is in reasonable agreement, but we have disagreement for the centers of gravity.

   Further gains to the accuracy reported here may  be made by using a Ramsey-Bord\'e interferometer~\cite{Bar1979, Rie1992, Cel1994, McF2009a, Fri2008}, where one should be able to reveal line shapes with the natural linewidth, but likely at the expense of \snr.   
   Our measurements may help to validate tests of many-body calculations applied to ytterbium~\cite{Dzu2011}. 
   There is also the potential to improve estimates on the nuclear charge radii variation  between isotopes and the specific mass shift using a King plot analysis as  described by Clark \textit{et al.},~\cite{Cla1979}, but by relying on these and other optical transition measurements in Yb, rather than muonic or x-ray shift data.   This seems pertinent given recent proposals  to use isotope shifts to search for phenomena beyond the Standard Model~\cite{Fru2017, Ber2018}. 
   

\section*{ }
\begin{acknowledgments}
This work was supported by the \ARC's Centre of Excellence for \EQUS\ (EQUS) (grant CE170100009).  J.~S. acknowledges support from the \UWA's   Winthrop Scholarship F74810.  P.A. was supported by an EQUS student scholarship.  We thank Jacinda Ginges and  Lilani Toms-Hardman for valuable discussions and Catriona Thomson for careful reading of the manuscript.   %
\end{acknowledgments}

\section*{Appendix}

In Table~\ref{SystShifts} we list relevant experimental parameters for each intercombination line. The mode number is that of the frequency comb mode  (the integer number of repetition rate spacings from zero frequency) to which the 1112\,nm laser frequency lies closest.  $f_\mathrm{DDS}$ is the frequency of the direct digital synthesiser that allows for changes to the stabilised repetition rate.  The systematic and statistical uncertainties \wrt\ the \Ybsix\ line are also listed. 
In Table~\ref{HFSummary3} we summarise the hyperfine energy shifts for the fermionic isotopes in terms of the respective hyperfine constants. 

\begin{table*} []
 \caption{Summary of the experimental parameters for \coolingT\ isotopic shifts and hyperfine splittings in ytterbium, along systematic and statistical uncertainties with respect to \Ybsix.} \label{SystShifts}
 \begin{ruledtabular}
 \begin{tabular}{llllll}
   \vspace{-0.3cm}  \\
 Transition    & Mode number, $n$ 	& $f_\mathrm{DDS}$ & $\Delta u_\mathrm{syst/176}$ & $\Delta u_\mathrm{stat/176}$\\
   (lower $\rightarrow$ upper)   &  $(+1078700)$ & $(+20\times10^6$\,Hz)  & (kHz)  & (kHz) \\
  \vspace{-0.3cm}  \\
 \hline
$^{173}$Yb\,($\frac{5}{2}\rightarrow\frac{7}{2}$)  & 68  & 380  & 31 & 16 \\		
$^{171}$Yb\,($\frac{1}{2}\rightarrow\frac{1}{2}$)  & 68  & 630  & 81 & 30 \\			
$^{176}$Yb & 71  &  250 &0  & 0 \\						
$^{174}$Yb & 73 & 170 & 16 & 25 \\			
$^{172}$Yb	& 75 & 170 &27 & 21\\	
$^{170}$Yb	& 77 & 480 & 26 & 29 \\
$^{173}$Yb\,($\frac{5}{2}\rightarrow\frac{5}{2}$) & 77 & 520 & 92 & 51 \\
$^{171}$Yb\,($\frac{1}{2}\rightarrow\frac{3}{2}$) & 80 & 510 &52 & 32\\
$^{173}$Yb\,($\frac{5}{2}\rightarrow\frac{3}{2}$) & 80 & 510 & 52& 57 \\  %
 \end{tabular}
 \end{ruledtabular}
 \end{table*}

\begin{table} [] \caption{Hyperfine energy shifts for \Yb\ and \Ybthree.  } \label{HFSummary3}
 \begin{ruledtabular}
 \begin{tabular}{lllll}
   \vspace{-0.3cm}  \\
Isotope &  $F$ & $E_{HF}/h$ \\
  \vspace{-0.3cm}  \\
 \hline
171 & 1/2 & $-A^{(171)}$ \\
 171 &  3/2 & $A^{(171)}/2$ \\
 173 &  3/2 &  $-7A^{(173)}/2+7 B^{(173)}/10$ \\
173 &  5/2 &  $-A^{(173)}-4 B^{(173)}/5$ \\
 173 &  7/2 &   $5A^{(173)}/2+ B^{(173)}/4$  \\
 \end{tabular}
 \end{ruledtabular}
 \end{table}
 


\end{document}